\documentclass[12pt]{article}
\usepackage{epsfig}
\def\makeatletter{\catcode`\@=11}
\makeatletter
\def\mathbox#1{\hbox{$\m@th#1$}}%
\def\math@ccstyles#1#2#3#4#5#6#7{{\leavevmode
      \setbox0\mathbox{#6#7}%
      \setbox2\mathbox{#4#5}%
      \dimen@ #3%
      \baselineskip\z@\lineskiplimit#1\lineskip\z@
      \vbox{\ialign{##\crcr
             \hfil \kern #2\box2 \hfil\crcr
             \noalign{\kern\dimen@}%
             \hfil\box0\hfil\crcr}}}}
\def\mathaccstyles{\math@ccstyles\maxdimen}
\def\maththroughstyles{\math@ccstyles{-\maxdimen}}
\def\unitmatrixDT%
 {\maththroughstyles{.45\ht0}\z@\displaystyle {\mathchar"006C}\displaystyle 1}

\def\unit {\unitmatrixDT}
\def\trace{{\rm tr\,}}
\def\tfrac#1#2{{\textstyle{#1\over #2}}}

\def\two{\tfrac{1}{2}}
\def\four{\tfrac{1}{4}}
\def\eight{\tfrac{1}{8}}
\def\sixteen{\tfrac{1}{16}}
\def\threesixteen{\tfrac{3}{16}}
\begin{document}
\renewcommand{\arraystretch}{1.5}

\rightline{UG-10/97}
\rightline{MRI-PHY/P970927}
\rightline{hep-th/9711160}
\vspace{3truecm}
\centerline{\bf Intersecting Branes in Matrix Theory}
\vspace{1cm}
\centerline{M.~de Roo, S.~Panda\footnote{Permanent Address:
 Mehta Research Institute of Mathematics \& Mathematical Physics,
 Chatnag Road, Jhusi, Allahabad 211506, India.}
 and J.~P.~van der Schaar}
\vspace{0.6truecm}
\centerline{\it Institute for Theoretical Physics}
\centerline{\it Nijenborgh 4, 9747 AG Groningen}
\centerline{\it The Netherlands}
\vspace{2truecm}
\centerline{ABSTRACT}
\vspace{.5truecm}
We construct BPS states in the matrix description of $M$-theory. 
 Starting from a set of basic $M$-theory branes, we study pair
 intersections which preserve supersymmetry. The fractions of
 the maximal supersymmetry obtained in this way are $\two,\ \four,\ 
 \eight,\ \threesixteen$ and $\sixteen$. In explicit examples we 
 establish that the matrix BPS states correspond to (intersecting)
 brane configurations that are obtained from the $d=11$ supersymmetry
 algebra. This correspondence for the 1/2 supersymmetric branes
 includes the precise relations between the charges.

\newpage

\noindent{\bf 1.\ Introduction}
\vspace{0.5cm}

\noindent Recently, the matrix model formulated in \cite{BFSS} for the
 microscopic description of M-theory \cite{Witten,Hull95} has drawn a 
 considerable amount of attention. This model may be taken as a quantum 
 mechanical framework for non-perturbative string theory (see 
 \cite{Banks} for a recent review). In this model the only degrees of 
 freedom are the zerobranes. However, various authors have successfully 
 demonstrated how the dynamics of strings, membranes and higher branes 
 can arise in this model.
 The matrix description of a membrane can be found in \cite{BFSS}, 
 while the open membrane is described in \cite{Li}. 
 A proposal for the description of a fourbrane (the wrapped fivebrane 
 of $M$-theory) is provided in \cite{BD}. 

In \cite{BSS} these higher dimensional objects are studied through the 
 supersymmetry algebra of the matrix description of $M$-theory. Thus, 
 the existence of conserved charges associated with the membrane
 and fivebrane is established. Interactions involving 
 different branes have been studied in, e.g., \cite{AB,LM,Lifs,che-tse}. 

In this note we provide additional evidence in support of the matrix model 
 from an investigation of matrix configurations that preserve some fraction of
 the maximal supersymmetry, and correspond to  
 intersecting branes. We start from a small number of basic objects with
 1/2 supersymmetry, which have nonzero 2-, 4-, 6- and 8-form charges.
 Besides these, some basic objects with less supersymmetry can be obtained.
 These configurations and their intersections should correspond to
 BPS solutions of the $d=11$ supergravity theory. 
 By an explicit analysis we establish this correspondence for the basic 
 objects and their pair intersections.

The fractions of maximal supersymmetry which can be obtained in this way
 are $\two,\ \four,\ \eight,\ \threesixteen$ and $\sixteen$. 
 In the next section we discuss the basic objects. Pair
 intersections, starting with an explicit construction of
 the matrix configuration corresponding to
 two intersecting membranes, are discussed in Section 3. 
 The correspondence of the matrix configurations to $d=11$ supergravity
 is discussed in Section 4 through an analysis of the $d=11$ supersymmetry 
 algebra.

\vspace{0.5cm}
\noindent{\bf 2.\ Basic solutions and residual supersymmetry}
\vspace{0.5cm}

\noindent The supersymmetric quantum mechanical theory which corresponds 
 to the matrix version of $M$-theory can be written, in a suitable 
 parametrisation \cite{dWHN,DFS,KP,BFSS}
\begin{equation}
\label{Mthlagr}
  L = {1\over 2g}\trace\bigg\{
   (\partial_0 X^a)^2 + 2 \theta_\alpha \partial_0\theta_\alpha
   - {1\over 2} \lbrack X^a,X^b \rbrack^2 - 2\theta_\alpha
   (\gamma_a)_{\alpha\beta}\lbrack\, \theta_\beta,X^a\rbrack \bigg\}\,.
\end{equation}
Here $a,b=1,\ldots,9$ correspond to the nine transverse 
 directions in the matrix model, $\alpha,\beta=1,\ldots,16$ 
 are nine-dimensional spinor indices\footnote{The 
 $\gamma$-matrices satisfy $\left\{\gamma_a,\gamma_b\right\}=2\delta_{ab}$.
 We use the notation $\gamma_{(n)}\equiv
 \gamma_{i_1\ldots i_n}=\gamma_{i_1}\gamma_{i_2}\cdots
 \gamma_{i_n}$. Note that $\gamma_{(n)}^2=1$ for $n=1,4,5,8,9$.},
 $X$ and $\theta_\alpha$ are hermitian $N\times N$-matrices. 
 It is understood that the limit $N\to \infty$ has to be taken, 
 although one can also give meaning to the finite $N$ models \cite{Susskind}.

The action (\ref{Mthlagr}) is invariant under the supersymmetry transformations
\begin{eqnarray}
  \delta X^a &=& -2\bar\epsilon \gamma^a\theta\,,
\nonumber\\
\label{susytrans}
  \delta \theta &=& {1\over 2}\bigg\{
    \partial_0 X^a\gamma_a + {i\over 2}\lbrack X^a,X^b\rbrack\gamma_{ab}
    \bigg\}\,\epsilon + \tilde\epsilon\,,
\end{eqnarray}
where $\epsilon$ and $\tilde\epsilon$ are independent supersymmetry 
 parameters. The algebra of supersymmetry transformations is given in 
 \cite{dWHN,BSS}, and contains besides the usual translational term 
 contributions of 2-form and 4-form charges\footnote{We normalize the
 charges $Z_{2n}$ with factors of $R_{11}^{n-1}$. For $Z_2$ and $Z_4$
 this is in agreement with \cite{BSS}. The factors in $Z_6$ and $Z_8$
 ensure that all charges have the same dimension. They are in agreement
 with the analysis of the $d=11$ supersymmetry algebra in Section 4.}
\begin{eqnarray}
  Z_2^{a_1a_2} &=& {i}\,\trace X^{[a_1}X^{a_2]}\,,\nonumber\\
  Z_4^{a_1a_2a_3a_4} &=& R_{11}\trace X^{[a_1}X^{a_2}X^{a_3}X^{a_4]}\,.
\end{eqnarray}
We also define 6- and 8-form charges:
\begin{eqnarray}
  Z_6^{a_1\cdots a_6} &=& iR_{11}^2\,\trace X^{[a_1}X^{a_2}\cdots X^{a_6]}\,,
\nonumber\\
  Z_8^{a_1\cdots a_8} &=& R_{11}^3\trace X^{[a_1}X^{a_2}\cdots X^{a_8]}\,.
\end{eqnarray}
Here $R_{11}$ is the radius of the compact direction $X^{11}$ in the
 matrix model. The momentum $P_{11}$ in that direction is given by
 $P_{11}=N/R_{11}$. Nonzero charges can only occur in the limit 
 $N\to \infty$. 

To see how objects with non-vanishing charges $Z_n$ can be constructed
 in matrix theory, we start
 with the single branes preserving $1/2$ of supersymmetry.
 
These are \cite{BSS}
\begin{itemize}
\item{$W$, the wave in the $a$-direction: 
  here we have $\partial_0(X^a)_{ij} = p^a\delta_{ij}$.}
\item{$M2$, the membrane: in this case the $X^a$ are time-independent, and
 for a membrane in the 12-direction we require 
 $\lbrack X^1,X^2\rbrack_{ij} = ic_{1}\delta_{ij}$, 
 where $c_1$ is a real parameter. To obtain a finite membrane charge
 $Z_2$, $c_1$ should scale as $N^{-1}$ for $N\to\infty$.}
\item{$M5$, the fivebrane which is wrapped around the longitudinal
direction. 
 For an $M5$ in the 1234-direction, we have 
 $\lbrack X^1,X^2\rbrack_{ij} = ic_{1}\delta_{ij},
\ \lbrack X^3,X^4\rbrack_{ij} = ic_{2}\delta_{ij}$.
 The charge of $M5$ is built up out of membrane charges. This object can be
 thought of as infinite stacks of membranes in both the 12-, and the 
 34-direction. Finite $Z_4$ again requires that the $c_i$ scale appropriately
 as $N\to\infty$.}
\item{$M6$, the sixbrane: in this case
$\lbrack X^1,X^2\rbrack_{ij} = ic_{1}\delta_{ij},
\ \lbrack X^3,X^4\rbrack_{ij} = ic_{2}\delta_{ij},\ 
 \lbrack X^5,X^6\rbrack_{ij} = ic_{3}\delta_{ij}$. 
 This is built up out of
 membranes in the 12-, 34-, and 56-directions, but there are also 
 non-vanishing fivebrane charges.
 Presumably $M6$ is
 related to the Kaluza-Klein monopole in $d=11$, although this 
 correspondence has not been established.}
\item{$M9$, the ninebrane, which is wrapped around the longitudinal
direction. Here
$\lbrack X^1,X^2\rbrack_{ij} = ic_{1}\delta_{ij},
\ \lbrack X^3,X^4\rbrack_{ij} = ic_{2}\delta_{ij},\ 
 \lbrack X^5,X^6\rbrack_{ij} = ic_{3}\delta_{ij},
\ \lbrack X^7,X^8\rbrack_{ij} = ic_{4}\delta_{ij}$.
 Again we have infinite stacks of membranes, as well as nonzero five- 
 and sixbrane charges.}
\end{itemize}

Since higher dimensional objects are built out of stacks of membranes,
 the charges $Z_{2n}$ can, for any $n$, be related to membrane
 charges. We find,
 independently of the choice of the scaling of 
 $\lbrack X^a,X^b \rbrack$, the behaviour
\begin{equation}
\label{Zrelation}
  Z_{2n} = P_{11}^{1-n} \prod_{i=1}^n Z_{2,i}\,\qquad n=1,\ldots,4\,.
\end{equation}
 The result (\ref{Zrelation}) is in agreement with the results of Section 4
 when considering 1/2 supersymmetric non-threshold states. 

These solutions to the matrix model equations of motion have $\theta=0$
 and preserve $1/2$ supersymmetry. 
 The vanishing of $\delta\theta$ (for static solutions, the 
 preservation of supersymmetry for $W$ is shown in a similar way) implies
that
\begin{equation}
\label{tildeeps}
  \delta_{ij}\tilde\epsilon = 
  - {i\over 4}\lbrack X^a,X^b\rbrack_{ij}\gamma_{ab}\epsilon \,,
\end{equation}
where the indices $i,j=1,\ldots,N$ have been made explicit.
 The relation (\ref{tildeeps}) can only be satisfied if 
\begin{equation}
\label{comm2}
 \lbrack X^a,X^b\rbrack_{ij} = iF^{ab}\delta_{ij}\,.
\end{equation}
A representation of (\ref{comm2}) can be given in terms of a
 pair of operators $p$ and $q$ satisfying canonical 
 commutation relations $\lbrack q,p\rbrack = i$. 
 As long as the commutator of the matrices $X$ is
 proportional to the unit matrix $\tilde\epsilon$ is determined
 in terms of $\epsilon$, so that 1/2 of supersymmetry is preserved.

Other basic solutions have less supersymmetry. We will consider the
 following ones:
\begin{itemize}
\item{$P5$, the pure fivebrane \cite{BSS}. This has the following structure:
 $\lbrack X^1,X^2\rbrack_{ij} = 
  \lbrack X^3,X^4\rbrack_{ij} = ic_{1}(\unit\otimes\sigma_{3})_{ij}$.
 We call it the pure fivebrane since the membrane charges vanish.
 Here 1/4 supersymmetry is preserved (see below).}
\item{$P9$, the ``pure'' ninebrane. Here we have
 $\lbrack X^1,X^2\rbrack_{ij} = ic_{1}(\unit\otimes\sigma_{3})_{ij},\ 
  \lbrack X^3,X^4\rbrack_{ij} = ic_{2}(\unit\otimes\sigma_{3})_{ij},\ 
  \lbrack X^5,X^6\rbrack_{ij} = ic_{3}(\unit\otimes\sigma_{3})_{ij},\ 
  \lbrack X^7,X^8\rbrack_{ij} = ic_{4}(\unit\otimes\sigma_{3})_{ij}$.
 This object is not entirely pure, since the constituent $P5$-charges
 do not vanish. However, there is no $M2$ or $M6$ charge.
 Depending on the values of the coefficients, $2n$, $n=1,2,3$
 of the 32 supersymmetry charges are unbroken (see below).}
\end{itemize}

\noindent Note that we cannot define a ``pure'' $M6$ ($P6$) in the same way,
 since then the charge of rank six vanishes. By using a more complicated 
 tensor structure for the matrices we can form a $P6$ and $P9$, 
 but these configurations do not preserve supersymmetry.

Let us now discuss the residual supersymmetry of the two solutions
 $P5$ and $P9$. We first consider $P9$. There are two equations
 that must have a solution to preserve some supersymmetry ($c_i\ne 0$),
 which imply $\tilde\epsilon=0$ and
\begin{equation}
\label{residualsusy}
   (c_1\gamma_{12}+c_2\gamma_{34}+c_3\gamma_{56}
    +c_4\gamma_{78})\,\epsilon = 0\,.
\end{equation}
We rewrite this  as $(1 - P)\,\epsilon=0$, with
\begin{equation}
 P = (c_2\gamma_{1234}+c_3\gamma_{1256}+c_4\gamma_{1278})/c_1\,.
\end{equation}
The $\gamma$-matrices in $P$ all square to one, and commute with each
 other. Also their trace, and the trace of their products, vanishes.
 These conditions determine the eigenvalues of
 $P$. Depending on the values of the coefficients, $2n$, $n=1,2,3$
 of the eigenvalues of $P$ can be equal to 1. 
 We find $n=1$, or preservation of 1/16 of the maximal supersymmetry,
 if, e.g., $c_1=\pm(c_2+c_3+c_4)$. For $n=2$, or 1/8, we need more
 stringent conditions: $c_1=c_2,\ c_3=c_4$. In that case the
 fivebrane charges in the directions 1234 and 5678 are still arbitrary
 (proportional to $c_1^2$ and $c_3^2$, respectively), but the
 other fivebrane charges (1256, 1278, 3456, 3478) are equal
 and proportional to $2c_1c_3$. The amount of preserved supersymmetry
 can be further increased by setting all coefficients equal:
 $c_1=c_2=c_3=c_4$. This corresponds to equal fivebrane charges in all 
 six directions, and 3/16 of the maximal supersymmetry.
 
If one of the coefficients, say $c_4$, 
 in (\ref{residualsusy}) vanishes, and we choose $c_1=\pm(c_2+c_3)$,
 then 1/8 supersymmetry is preserved. 
 This can be interpreted as an intersection of oppositely charged
 sixbranes, a configuration which also has nonzero fivebrane charges.
 If two coefficients vanish, the remaining two 
 must be equal to preserve 1/4 supersymmetry. 
 This last case corresponds to $P5$.

This supersymmetry analysis is
 very similar to that occurring in the 
 analysis of branes which intersect at angles \cite{PKT,OT,jab}.

\vspace{0.5cm}
\noindent{\bf 3.\ Pair intersections}
\vspace{0.5cm}

The fact that for $P5$ and $P9$ the commutators of the $X^a$ are
 not proportional to the unit matrix is the cause of the additional
 supersymmetry breaking. For intersecting
 pairs we split the matrices in two blocks, each representing a brane,
 of size $N_1$ and $N_2$, with $N_1+N_2=N$.
 We will limit ourselves in this paper to pair intersections,
 starting with those involving the wave $W$.

It is easy to see that only in the case of the branes
 $M2$, $M5$, $M6$ and $M9$ a supersymmetric intersection with a wave can be 
 constructed\footnote{For $P5$ and $P9$ one finds
 the requirement $\gamma_1\epsilon=0$ for a wave in the 1-direction (since
 $\tilde\epsilon$ vanishes). However, $\gamma_1$ has no zero eigenvalues.}.
 In these cases the direction of the wave necessarily must be in the
 worldvolume of the brane\footnote{Consider a membrane in the 12-direction.
 If the wave is not in the worldvolume
 of the brane, the condition on $\epsilon$ is of the form
 $(c_1\gamma_{12} - p\gamma_9)\,\epsilon=0$ for a wave in the 9-direction.
 The $\gamma$ matrices can be simultaneously diagonalised. Since
 $\gamma_9$ has real, and $\gamma_{12}$ imaginary eigenvalues, their
 linear combination cannot have eigenvalue zero. For branes of higher 
 dimension the same argument holds.}.
 The same analysis we did in Section 2 reveals
 that the possible fractions are 1/4, 1/8, 3/16 and 1/16. This case is
 summarized in Table 1.

Let us now look at the pair intersections of $M2$, $M5$, $M6$ and $M9$.
 In the general case, the condition (\ref{residualsusy}) will be of the
 form
\begin{equation}
      F^{ab}\gamma_{ab}\epsilon=0\,,\qquad F^{ab}\equiv F^{ab}_1-F^{ab}_2\,,
\end{equation}

\begin{eqnarray*}
\begin{array}{|c||c|c|}
\hline
\mbox{Matrix configuration}&\mbox{SUSY}\\
\hline
\hline
(1|W,M2)  & \tfrac{1}{4}  \\
(1|W,M5)  & \tfrac{1}{4}  \\
(1|W,M6)  & \tfrac{1}{8}  \\
(1|W,M9)  & \eight,\ \tfrac{1}{16},\ \tfrac{3}{16}  \\
\hline
\end{array}\nonumber
\end{eqnarray*}

\vspace{3mm}
{\scriptsize
\noindent Table 1.\ {\bf Supersymmetric pair intersections involving $W$.} 
 The notation $(p|A,B)$ indicates that the objects $A$ and $B$ have
 $p$ common spacelike worldvolume directions. The second column gives the
 amount of residual supersymmetry that can be obtained.
}
\vspace{5mm}

\begin{eqnarray*}
\begin{array}{|c|c||c|c|}
\hline
\mbox{Configuration}&\mbox{SUSY}&\mbox{Configuration}&\mbox{SUSY}\\
\hline
\hline
(0|M2,M2) & \four         
  &(4|M5,M5) & \two,\ \four  \\
(2|M2,M2) & \two          
  &(2|M5,M6) & \eight,\ \sixteen,\ \threesixteen \\
(0|M2,M5) & \eight        
  &(4|M5,M6) & \four,\ \eight \\
(2|M2,M5) & \four         
  &(4|M5,M9) & \four,\ \eight,\ \sixteen,\ \threesixteen \\
(0|M2,M6) & \eight,\ \sixteen,\ \threesixteen 
  &(4|M6,M6) & \four,\ \eight,\ \sixteen,\ \threesixteen \\
(2|M2,M6) & \four,\ \eight  
  &(6|M6,M6) & \two,\ \four,\ \eight  \\
(2|M2,M9) & \eight.\ \sixteen,\ \threesixteen  
  &(6|M6,M9) & \four,\ \eight,\ \sixteen,\ \threesixteen \\
(0|M5,M5)   & \eight,\ \sixteen,\ \threesixteen 
  &(8|M9,M9) & \two,\ \four,\ \eight,\ \sixteen,\ \threesixteen  \\
(2|M5,M5) & \four,\ \eight  
  & \hfil & \hfil \\
\hline
\end{array}\nonumber
\end{eqnarray*}

\vspace{3mm}
{\scriptsize
\noindent Table 2.\ {\bf Pair intersections of $M2$, $M5$, $M6$, $M9$.} 
 Only branes are considered which are built up out of membranes in the
 12, 34, 56 and 78 directions.
}
\vspace{5mm}

\noindent where 
 $F_i$, $i=1,2$ come from the commutators $\lbrack X^a,X^b\rbrack$
 for the two branes. Using nine-dimensional rotations a generic
 antisymmetric matrix can be put into a canonical form, in which
 only $F^{12},\ F^{34},\ F^{56}$ and $F^{78}$ are nonzero. Thus the
 analysis reduces to that of (\ref{residualsusy}). Therefore
 the only fractions of maximal supersymmetry\footnote{Note
 that in \cite{OT} also fractions $\tfrac{1}{32},\ \tfrac{3}{32}$ 
 and $\tfrac{5}{32}$ 
 are obtained for pair intersections at angles, but these require
 ten spatial dimensions.
 For orthogonal intersection of branes also
 $\tfrac{1}{32}$ can occur \cite{BdREJS}, 
 but this requires at least five branes.} 
 in pair intersections are
 $\tfrac{1}{4},\ \tfrac{1}{8},\ \tfrac{1}{16},\ \tfrac{3}{16}$.

We will limit ourselves to those cases for which the only nonzero
 commutators used in constructing the branes are 
 $\lbrack X^{2n-1},X^{2n}\rbrack$ for $n=1,\ldots,4$. For such configurations
 the pair intersections are summarised in Table 2. 
 As an illustration
 we will work out one particular case, the intersection of a membrane
 $M2$ with the $M6$-brane, in detail.

 Splitting up the matrices appropriately and using 
(\ref{susytrans}) we get the following equations for the supersymmetry 
parameters
\begin{eqnarray}
M6: \hspace{2cm} &\tilde\epsilon=&(c_1\gamma_{12}+c_2\gamma_{34}+
         c_3\gamma_{56})
\epsilon\,, \nonumber \\
M2: \hspace{2cm} &\tilde\epsilon=&c_4 \, \gamma_{12} \epsilon\,.
\end{eqnarray}
The first equation breaks half of the supersymmetry and for the second 
equation to be consistent with the first we find that
\begin{equation} 
    ((c_1-c_4)\gamma_{12}+c_2\gamma_{34}+c_3\gamma_{56})\epsilon =0\,.
\end{equation}
When $c_1=c_4$ we must have $c_2=\pm c_3$, and 1/4 supersymmetry is preserved.
 If $c_1\ne c_4$, we must require $c_2+c_3=c_1-c_4$ (up to choices of signs)
 to preserve 1/8 of the maximal supersymmetry.

We can also have $(0|M2,M6)$, with the membrane directions
 outside the $M6$. This leads to equation
 (\ref{residualsusy}), and can preserve $n/16$, $n=1,2,3$ of the
 maximal supersymmetry.

When one brane in the pair is a `pure' brane ($P5$ or $P9$) the analysis 
 changes. Because a `pure' brane makes $\tilde\epsilon=0$, for every
 brane in the pair we get an equation $R\,\epsilon=0$ where $R$ is
 the sum of one or more matrices $\gamma_{(2)}$.
 So we have to look for zero eigenvalues of the 
 matrix $R$. 
 This means that we cannot add an $M2$ or $W$ to a 
 $P5$ or $P9$, because in those cases $R$ has no zero eigenvalues. 
 The 
 preserved supersymmetry depends on the relative orientation and on the number 
 of $\gamma$-matrices in each $R$.  The fractions of
 supersymmetry that can be obtained are the same as in the cases
 considered previously.

\vspace{0.5cm}
\noindent{\bf 4.\ Relation with eleven-dimensional supergravity}
\vspace{0.5cm}

The supersymmetry algebra in $d=11$, including all allowed central charges,
 takes on the following form \cite{vHvP,PKT2}:
\begin{equation}
\label{D11QQ}
 \{Q_\alpha,Q_\beta\} = 
 (C\Gamma^m)_{\alpha\beta}P_m +
 \two(C\Gamma^{mn})_{\alpha\beta} Z_{mn} +
 \tfrac{1}{5!}(C\Gamma^{m_1\ldots m_5})_{\alpha\beta}Z_{m_1\ldots m_5}\,,
\end{equation}
where $\alpha, \beta=1,\ldots,32$.
The charges $Z_{mn}$ and $Z_{m_1\ldots m_5}$ correspond, 
 in the case of spacelike indices, to the membrane charge and
 fivebrane charge. If one of the indices is timelike, $Z_{0m}$
 and $Z_{0m_1\ldots m_4}$ correspond to the dual of a ninebrane 
 and a sixbrane charge, respectively \cite{Hull97}. 
 We choose $C=\Gamma^0$ and write
\begin{equation}
  \{Q,Q\} = P^0(\unit +\bar\Gamma)\,.
\end{equation}

In matrix theory in the infinite momentum frame there is always a wave
 present, which, in this section, we place in the direction 9. 
 The basic $M5$ configuration
 corresponds to nonzero $P_9=p$, $Z_{12}=z_1$ and $Z_{34}=z_2$
 (because $M5$ has nonzero membrane charges) and $Z_{12349}=y$, 
 with of course a component in the direction of the boost. We find
\begin{equation}
\label{D11M5}
  (\bar\Gamma)^2 = (P^0)^{-2}\left(
       p^2+z_1^2+z_2^2+y^2 + 2(py-z_1z_2)\Gamma^{1234}\right)
\end{equation}
If the charges are such that $py=z_1z_2$ then we can choose $P^0$
 to set $(\bar\Gamma)^2=\unit$, which implies that 1/2 of the maximal
 supersymmetry is preserved. 
 This relation between the momentum and the charges is  what
 we expect from the matrix theory ((\ref{Zrelation}) for $n=2$).

The pure fivebrane, $P5$, has no membrane charges, and therefore
\begin{equation}
\label{D11P5}
  (\bar\Gamma)^2 = (P^0)^{-2}\left(
       p^2 + y^2 + 2py\Gamma^{1234}\right)\,.
\end{equation}
Now we cannot set $(\bar\Gamma)^2=\unit$, but we can set 16 of the eigenvalues
 of $(\bar\Gamma)^2$ equal to one by choosing $P^0$ appropriately.
 This means that $\bar\Gamma$ has 8 eigenvalues equal to $-1$,
 and $1/4$ supersymmetry is unbroken. This $d=11$ configuration corresponds 
 to a fivebrane and a wave.

In this way the matrix configurations of Section 2 can be identified with
 supergravity solutions. $M5$ corresponds to a bound state of
 two membranes and a fivebrane, boosted in the 9 direction
 (see also the discussion in \cite{sorto}). With 1/2
 supersymmetry this is a non-threshold solution, which is not yet known
 as a solution of the $d=11$ supergravity equations. The
 result (\ref{D11P5}) for $P5$ corresponds to a threshold solution,
 and is the known intersection of a fivebrane and a wave \cite{Tseytlin95}.

To find corresponding BPS states for the 1/2 supersymmetric matrix $M6$ and
$M9$ the same analysis can be done as for the $M5$. The result is that these 
(non-threshold) states do exist in the supersymmetry algebra, but we have to 
impose constraints on the charges. These constraints however are exactly  
the relations between the different charges (\ref{Zrelation}) in  matrix 
theory.

As is clear from Table 2, there are configurations preserving 3/16
 of the maximal supersymmetry. The case of $(0|M5,M5)$\footnote{This 
 configuration corresponds to two fivebranes intersecting over a string. In
 matrix theory the common direction corresponds to the longitudinal 
 direction and that is why we write $(0|M5,M5)$.} was studied in detail
 in \cite{ohzhou}. These authors show that this configuration in $d=10$ is
 T-dual to two $D4$ branes at angles.
  
As an example of a state preserving 3/16 of the supersymmetry we analyse
 $P9$. There is
 one ninebrane charge, mixed with 6 fivebrane charges and 
 momentum in the 9th direction. The ninebrane charge corresponds 
 to\footnote{$\natural$ indicates the direction 10. Note that 
 $\Gamma^{\natural} = \Gamma^{0123456789}$.}
 $Z_{0\natural}=m$. Including all charges we obtain
\begin{eqnarray}
 P^0\bar\Gamma &=& \Gamma^{09}p +\Gamma^{012349}y_1 
   +\Gamma^{012569}y_2 +\Gamma^{012789}y_3  \nonumber\\
 &&\qquad   + \Gamma^{034569}y_4 +\Gamma^{034789}y_5 
   +\Gamma^{056789}y_6 +\Gamma^{\natural}m \,,
\end{eqnarray}    
In $(P^0\bar\Gamma)^2$
 there are three independent commuting $\Gamma$-matrices 
 so that in the generic case
 this configuration will preserve 1/16 of the supersymmetry.
 This corresponds to a threshold bound state of six fivebranes, a ninebrane
 and a wave. 
 We can also obtain configurations 
 which preserve 1/8 and 3/16, by restricting the
 coefficients. If we set $y_2=y_3=y_4=y_5=y$, leaving
 $y_1$ and $y_6$ arbitrary, we find that
 $(P^0\bar\Gamma)^2$ has the following eigenvalues:
 $(p-m\pm(y_1-y_6))^2$ with multiplicity 8 for each choice of sign,
 $(p+m+y_1+y_6)^2$ with multiplicity 8, and
 $(p+m-y_1-y_6\pm 4y)^2$ with multiplicity 4 for each sign. Therefore,
 by choosing $P^0$ appropriately, we preserve 1/8 
 supersymmetry, for each of the eigenvalues of multiplicity 8.
 If we also set $y_1=y_6=y$, the eigenvalues simplify further. There are then
 12 eigenvalues equal to $(p+m+2y)^2$, leading to 3/16 of the maximal
 supersymmetry.
 For six equal charges we find that
 for $(P^0)^2= (p-m)^2$ then 1/4 of the supersymmetry charges
 are preserved. Thus the $d=11$ supersymmetry
 algebra seems to support a boosted longitudinal ninebrane
 with 1/4 supersymmetry. In Section 2 we showed that such an object is
 absent in the matrix model.

We believe that the supersymmetric configurations in matrix theory 
 that we constructed in Section 2 all correspond to supersymmetric
 states in the $d=11$ supersymmetry algebra (\ref{D11QQ}).
 We have verified this in a number of cases, and always found agreement. 
 Presumably, solutions of the $d=11$ supergravity equations of
 motion for such states can be constructed. A lot of work
 has been done on non-threshold states
 involving membranes and fivebranes (see for instance \cite{Russo,Ohta97}).
 In the case of the sixbrane or Kaluza-Klein monopole much less is known, 
 while of course the status of the ninebrane as a solution in $d=11$
 supergravity is uncertain.

However, not all supersymmetric configurations constructed
 in the $d=11$ supersymmetry algebra can be obtained from the matrix
 model.
 For instance, in the $d=11$ algebra
 the sixbrane together with a transverse wave gives a state with 16
 preserved supersymmetry charges\footnote{In this case we
 have $\bar\Gamma = (\Gamma^{09}p+\Gamma^{0123456}m)/P^0$, which
 corresponds to $\bar\Gamma^2 = (p^2+m^2)/(P^0)^2$.}. This we
 do not find in the matrix model. In the analysis of $P9$ given above the
 result in the $d=11$ algebra suggests a pure ninebrane with 1/4 
 supersymmetry in the matrix model. This is also absent in Section 2.

So, concerning the basic branes there seems to be a problem involving the 
 absence of pure sixbranes and ninebranes. In the matrix model
 they can be constructed but 
 break all of the supersymmetries, while the $d=11$ supersymmetry
 algebra seems to support supersymmetric configurations of this type.

\vspace{0.5cm}
\noindent{\bf 5.\ Conclusion}
\vspace{0.5cm}

Although the missing transversal fivebrane, as well as the problems
 involving six- and ninebranes, 
 indicate that something is still poorly understood in the matrix model,  
 many BPS states and their pair intersections
 seem to be in agreement with what we expect if 
 the matrix model is to describe $M$ theory.
 We believe that any matrix BPS state has an 
 analogue as a threshold or non-threshold intersecting
 brane configuration  in the $d=11$ supersymmetry algebra. 
 
In this paper we establish a (partial) correspondence between supersymmetric
 branes  in the matrix model and in the $d=11$ supersymmetry algebra.
 Especially, the 1/2 supersymmetric basic $M6$ and $M9$ in matrix theory 
 correspond exactly to 1/2 supersymmetric non-threshold states in the 
 supersymmetry algebra carrying the same charges. 
 Interesting open questions concerning the existence of non-threshold
 solutions to the $d=11$ supergravity equations of motion corresponding
 to matrix model states remain. The relations
 presented here
 between matrix theory and the supergravity limit can be considered
 additional evidence for matrix theory. 
 The fact that the correspondence is not complete implies that further
 work needs to be done, and hopefully this will lead to a better understanding 
 of matrix theory.

\vspace{0.5cm}
\noindent{\bf Acknowledgements}
\vspace{0.5cm}

This work is part of the Research program of the
``Stichting voor Fundamenteel Onderzoek der Materie''(FOM).
It is also supported by the European Commission TMR programme
ERBFMRX-CT96-0045, in which M.~de R. is associated
to the University of Utrecht. S.~P. is grateful to the group in Groningen
for their hospitality and support which made this collaboration possible.



\begin{thebibliography}{30}

\bibitem{BFSS}
T.~Banks, W.~Fischler, S.~H.~Shenker and L.~Susskind,
Phys.~Rev.~{\bf D55} (1997) 5112
{\tt hep-th/9610043}

\bibitem{Witten}
E. Witten, 
Nucl. Phys.~{\bf B443} (1995) 85,
{\tt hep-th/9503124}

\bibitem{Hull95}
C. Hull and P.K. Townsend, 
Nucl. Phys.~{\bf B438} (1995) 109,
{\tt hep-th/9410167}

\bibitem{Banks}
T.~Banks,
{\sl Matrix Theory},
Trieste Spring School on Supergravity and Superstrings, 1997,
{\tt hep-th/9710231}

\bibitem{Li}
M. Li, 
Phys.~Lett.~{\bf B397} (1997) 37,
{\tt hep-th/9612144}

\bibitem{BD}
M. Berkooz and M.R. Douglas, 
Phys.~Lett.~{\bf B395} (1997) 196,
{\tt hep-th/9610236}

\bibitem{BSS}
T.~Banks, N.~Seiberg and S.~H.~Shenker,
Nucl.~Phys.~{\bf B490} (1997) 91,
{\tt hep-th/9612157}

\bibitem{AB}
O. Aharony and M. Berkooz, 
Nucl.~Phys.~{\bf B491} (1997) 184,
{\tt hep-th/9611215.}

\bibitem{LM}
G. Lifschytz and S. D. Mathur, 
{\sl Supersymmetry and Membrane Interactions in M(atrix) Theory},
{\tt hep-th/9612087.}

\bibitem{Lifs}
G. Lifschytz, 
{\sl Four-Brane and Six-Brane Interactions in M(atrix) Theory},
{\tt hep-th/9612223.}

\bibitem{che-tse}
I.~Chepelev and A.~A.~Tseytlin,
{\sl Long-distance interactions of branes: correspondence between supergravity
and super Yang-Mills descriptions}, 
{\tt hep-th/9709087}

\bibitem{dWHN}
B.~de Wit, J.~Hoppe and H.~Nicolai,
Nucl.~Phys.~{\bf B305[FS23]} (1988) 545

\bibitem{DFS}
U.~H.~Danielsson, G.~Ferretti and B.~Sundborg,
Int.~Journ.~Mod.~Phys.~{\bf A11} (1996) 5463,
{\tt hep-th/9603081}

\bibitem{KP}
D.~Kabat and P.~Pouliot,
Phys.~Rev.~Lett.~{\bf 77} (1996) 1004,
{\tt hep-th/9603127}

\bibitem{Susskind}
L.~Susskind,
{\sl Another conjecture about M(atrix) theory},
{\tt hep-th/9704080}


\bibitem{PKT}
P.~K.~Townsend,
{\sl M-branes at angles},
{\tt hep-th/9708074}

\bibitem{OT}
N.~Ohta and P.~K.~Townsend, 
{\sl Supersymmetry of M-branes at angles},
{\tt hep-th/9710129}

\bibitem{jab}
M.~M.~Sheikh~Jabbari,
{\sl Classification of Different Branes at Angles},
{\tt hep-th/9710121}

\bibitem{BdREJS}
E.~Bergshoeff, M.~de Roo, E.~Eyras, B.~Janssen and J.~P.~van der Schaar,
Nucl.~Phys.~{\bf B494} (1997) 119,
{\tt hep-th/9612095}


\bibitem{vHvP}
J.~W.~van Holten and A.~Van Proeyen,
J.~Phys.~{\bf A15} (1982) 3763

\bibitem{PKT2}
P.~K.~Townsend,
{\sl p-brane democracy},
 in {\sl Particles, Strings and Cosmology}, eds.~J.~Bagger, G.~Domokos, 
 A.~Falk and S.~Kovesi-Domokos (World Scietific 1991),
{\tt hep-th/9507048}

\bibitem{Hull97}
C.~M.~Hull,
{\sl Gravitational Duality, Branes and Charges},
{\tt hep-th/9705162}

\bibitem{sorto}
D.~Sorokin and P.~K.~Townsend,
{\sl M-theory superalgebra from the M-5-brane},
{\tt hep-th/9708003}

\bibitem{Tseytlin95}
A.~Tseytlin,
Nucl.~Phys.~B475 (1996) 179,
{\tt hep-th/9604035}

\bibitem{ohzhou}
N.~Ohta and J.~G.~Zhou,
{\sl Realization of D4-Branes at Angles in Super Yang-Mills Theory},
{\tt hep-th/9709065}

\bibitem{Russo}
J.~G.~Russo and A.~A.~Tseytlin,
Nucl.~Phys.~{\bf B490} (1997) 121,
{\tt hep-th/9611047}

\bibitem{Ohta97}
N.~Ohta and J.-G.~Zhou,
{\sl Towards the classification of Non-Marginal Bound States of $M$-branes
 and their Construction Rules},
{\tt hep-th/9706153}

\bibitem{Gauntlett97}
J.~P.~Gauntlett, G.~W.~Gibbons, G.~Papadopoulos and P.~K.~Townsend,
Nucl.~Phys.~B500 (1997) 133,
{\tt hep-th/9702202}

\end{thebibliography}
\end{document}